\documentclass{aa}
\usepackage{graphics}
\begin{document} 


\def\ra{\rightarrow}
\def\deg{\ifmmode^\circ\else$^\circ$\fi}
\def\kms{km\thinspace s$^{-1}$}


\title{Cometary molecular clouds around RNO\,6} 
\subtitle{On-going star formation near the double cluster 
{\it h} and $\chi$ Persei}
\author{R. Bachiller, A. Fuente, and M.S.N. Kumar}

\institute{IGN Observatorio Astron\'{o}mico Nacional, Apartado 1143,
E-28800 Alcal\'{a} de Henares, Spain} 

\offprints{R.Bachiller \\ e-mail:bachiller@oan.es}
\date{Received: ; accepted: to be inserted later}
\authorrunning{Bachiller et al.}
\titlerunning{Star formation around RNO6}

\abstract{We present molecular line observations of the star-forming
cloud around RNO\,6 along with a newly discovered nearby molecular
cloud that we name RNO\,6\,NW. Both clouds display striking
similarities in their cometary structures and overall kinematics.  By
using $^{13}$CO line observations, we estimate that these clouds
have  similar sizes ($\sim$4.5\,pc) and masses ($\sim$200
M$_{\sun}$).\\
Both molecular clouds RNO\,6 and RNO\,6\,NW are active
in star formation. From new high resolution near-IR narrowband
images, we confirm that RNO\,6 hosts an embedded IR cluster that
includes a Herbig Be star. A conspicuous H$_2$ filament is found to
delineate the dense cometary head of the globule.  RNO\,6\,NW hosts
at least two IR sources and a bipolar molecular outflow of $\sim$0.9\,pc
of length and $\sim$0.5\,M$_{\sun}$ of mass.\\
We show that the
cometary structure of both clouds has been created by the UV
radiation from numerous OB stars lying $\sim$1.5{\deg} to the north. 
Such OB stars are associated with the double cluster $h$ and $\chi$
Persei, and are probably members of the Per\,OB1 association. Thus
star formation inside these clouds has been very likely triggered by
the Radiation Driven Implosion (RDI) mechanism. From comparison to
RDI theoretical models, we find that the similar kinematics and
morphology of both clouds is well explained if they are at a
re-expansion phase.  Triggered sequential star formation also
explains the observed spatial distribution of the members of the
near-IR cluster inside the RNO\,6 cloud, and the morphology of the
H$_2$ filament. We conclude that the RNO\,6 and RNO\,6\,NW clouds are
high-mass counterparts to the cometary globules of smaller masses
which have been studied up to now. Thus our observations demonstrate
that the RDI mechanism can produce, not only  low mass stars in small
globules, but also intermediate mass stars and clusters in massive
clouds. 
}

\maketitle

\section{Introduction} 

Red Nebulous Object 6 (RNO\,6, Cohen \cite{coh80}) is a rather bright
nebulosity of $\sim$1{\arcmin} size lying at the eastern border of an
optical extinction patch of $\sim$2{\arcmin} and placed $\sim$5$\deg$
below the galactic plane in the Perseus constellation. The object was
first catalogued as GM\,4 by Gyulbudagyan \& Magakyan (\cite{gm77}).
It was also included, with the name NS\,3, in the catalog of bipolar
nebulae by Neckel \& Staude (\cite{ns84}), because of the peculiar
morphology which includes an equatorial lane of obscuration, roughly
in the E-W direction, suggestive of a dusty disk. However,
polarization measurements by Scarrott et al. (\cite{scrt86}) have
shown that the dust obscuration in this lane is too low to constitute
a dense disk as those which are usually observed around protostars or
very young stars.

\begin{figure*}

\vskip -3cm
\hspace{-1cm}
\includegraphics{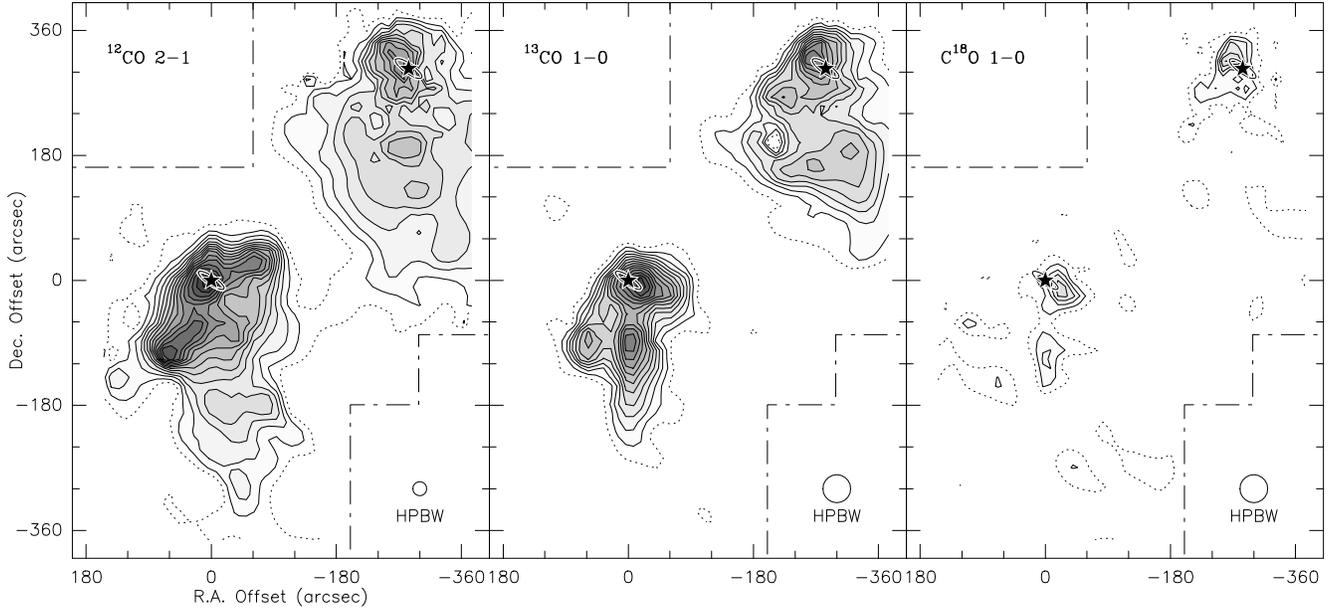}
\vskip 12cm
\caption{CO J=2$\rightarrow$1,   $^{13}$CO J=1$\rightarrow$0, and
C$^{18}$O  J=1$\rightarrow$0 integrated intensity maps of the RNO\,6
area. First contour and step are 3.5, 0.88, and 0.28  K \kms, for CO,
$^{13}$CO, and C$^{18}$O, respectively. The first contour is drawn
with dashed lines.  The dashed-dotted lines indicate the limits of
the mapped area. The star symbols surrounded by small ellipses 
indicate the nominal position of the two IRAS sources
IRAS\,02130+5509 and IRAS\,02124+5514 with their position
uncertainties. The telescope beam  is also indicated at the
right-bottom corner of each panel. Position offsets are with respect
to the nominal position of the IRAS\,02130+5509 source at 02:13:03.3,
+55:09:12 (1950.0). } 

\label{fig:1} 
\end{figure*}

RNO\,6 contains a B star  with H$\alpha$ emission (Cohen
\cite{coh80}) which thus fulfills all the criteria that define  
Herbig AeBe stars as a class, namely (i) its spectral type is A or
earlier with emission lines in the spectrum, (ii) its location is in
an obscured region, and (iii) it is illuminating a nebulosity. This
star has been hence included in some standard lists of HAeBe stars
(e.g. Th\'e et al. \cite{the94}). The systematic searches for
clustering around HAeBe stars carried out by Testi el al.
(\cite{tes97}, \cite{tes98}, \cite{tes99}) revealed a clear
enhancement in the stellar surface density profile toward the central
RNO\,6 position; it appears that the central B star is accompanied by
a small group of less luminous stars.

The distance to RNO\,6 is poorly determined.  The kinematic estimate
from a standard galactic curve (e.g. Burton \cite{bur74}) is 3$\pm$1
kpc, whereas the polarization measurements of Scarrott et al.
(\cite{scrt86}) together with photometric considerations indicate a
distance in the range 1.6 to 2.2 kpc. So we will assume in this paper
a distance of 2 kpc.  The spectral type of the main star of the
RNO\,6 group is also poorly known. The IRAS luminosity
($\sim$300\,$L_{\odot}$) corresponds to a ZAMS B6 star or to a B8III.
No radio continuum emission at $\lambda$ 6 cm was detected in the
sensitive VLA search by White \& Gee (\cite{wg86}).

The environment of RNO\,6 was essentially unstudied up to now. The
region was first observed in the millimeter wave range in the context
of a complete survey of HAeBe stars (Fuente et al. \cite{fuente01}),
and was found to be interesting on its own. One of the interests of 
the region resides in its possible relationship with the double
cluster $h$ and $\chi$ Persei,  which lie at only $\sim$1.5{\deg} in
the sky from RNO\,6, and which are at a comparable distance from the
Sun (2.2-- 2.6 kpc, according to Tapia et al. \cite{tap84}). Thus, as
an extension to the surveys by Fuente et al. (\cite{fuente01}), we 
undertook a detailed study of the RNO\,6 area, and the corresponding
results are presented in this paper. The new data provide important
information about the mass, kinematics, and evolutionary stage of the
region; we have found that the strong UV field from the OB
stars associated with the double cluster is creating  striking cometary
structures and is triggering star formation activity in molecular
clouds around RNO\,6.

\section{Observations} 

We observed the $^{13}$CO J=1$\rightarrow$0,  C$^{18}$O
J=1$\rightarrow$0, and $^{12}$CO J=2$\rightarrow$1 rotational
transitions  around RNO\,6 with the IRAM 30\,m radiotelescope  at
Pico Veleta (near Granada, Spain) in June 1998. The three transitions
were observed simultaneously using the multireceiver capabilities of
the 30\,m telescope.  The backend was an autocorrelator split in
several parts which provided  a spectral resolution $\sim$ 78 kHz.
Forward efficiency, main beam efficiency, typical system temperatures
and Half Power Beam Width were 0.92, 0.68, 350 K and 24$\arcsec$ at the
frequency of the $^{13}$CO J=1$\rightarrow$0 and C$^{18}$O
J=1$\rightarrow$0 lines, and 0.86, 0.39, 1000 K and 12$\arcsec$ at that of
the $^{12}$CO J=2$\rightarrow$1 line. Some regions of interest around
the C$^{18}$O maxima  and around newly-detected $^{12}$CO wings (see
Section 3) were explored in the H$^{13}$CO$^+$ J=1$\rightarrow$0, and
SiO J=2$\rightarrow$1 lines near $\lambda$ 3\,mm. The characteristics
of the telescope at these frequencies are similar to those at the
C$^{18}$O J=1$\rightarrow$0 frequency. No SiO emission was detected
at a level of 0.02 K r.m.s. (1.6 {\kms} velocity resolution). 
H$^{13}$CO$^+$ emission was well detected around the C$^{18}$O
maxima, but the data quality was not good enough as to produce
contour maps. All line intensities in this paper are reported in
units of main beam brightness temperature.

Near infrared (IR) observations were carried out with the United
Kingdom Infrared Telescope (UKIRT) at Mauna Kea (Hawaii, USA) as part
of a  service observing program on October 2000. Narrow band 
($\Delta\lambda =0.02 \mu$m) images through H$_{2}$
($\lambda=2.122\mu$m) and continuum ($\lambda=2.104 \mu$m) filters
were obtained using the near-IR camera UKIRT Fast Track Imager
(UFTI). UFTI employs a $1K\times1K$ HgCdTe array; the optics used
provides a pixel scale of 0.0917{\arcsec} and a field of view $\sim$
90\arcsec. Two positions centered on RNO\,6 and RNO\,6\,NW were
observed with a 3$\times$3 mosaic to cover a total field of view of
136\arcsec. As a result of the mosaic technique the signal to noise
ratio is higher in the central regions of the  final images than at
the edges. The average seeing during these observations was less than
0.5$\arcsec$.

\section{Overall structure} 

\subsection{Molecular clouds} 

We mapped a region of 10{\arcmin}$\times$13{\arcmin} around the
central RNO\,6 position in CO rotational lines. An overall view of
the molecular clouds in the area can be obtained from the integrated
intensity maps of Fig. 1. Strong CO emission around RNO6 comes from 
a molecular cloud which extends $\sim$7.5{\arcmin}$\times$4{\arcmin}.
Surprisingly, a second cloud of similar size is observed toward the
NW. In the following we will refer to these large-scale clouds as
RNO\,6 and RNO\,6\,NW, respectively. The central velocities of the
emission peaks in both clouds are very close ($\sim\,-36$ {\kms}).
Such a coincidence in radial velocity,  together with the ridge of
weak emission extending between both clouds, suggests that these
clouds are placed at similar distances from the Sun, and very likely
are physically connected. 

\begin{figure}
\resizebox{\hsize}{!}{\includegraphics{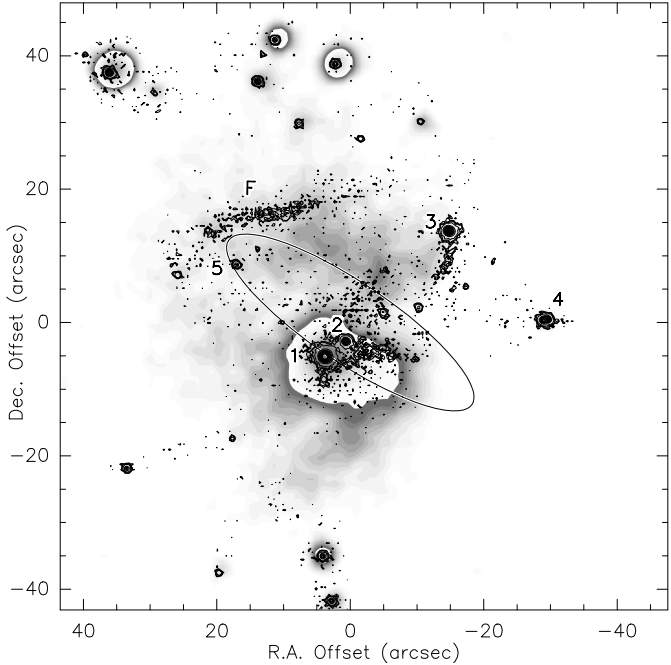}}

\caption{H$_2$ image contours of RNO\,6 overlayed on the red Digital
Sky Survey (DSS) image. The DSS picture is shown grey-scaled, but
note that the grey scale saturates, and starts again from white,
around the maxima in the brightest stars and near the center of the
image. This scaling is convenient to display near-IR features.
Continuum has not been subtracted in the near-IR image in order to
display the stars and the relative positioning of both images. Some
important features, as the relatively bright H$_2$ filament labeled
{\bf F}, and near-IR stars labeled with numbers are discussed in the
text. The ellipse corresponds to the positional uncertainty of
IRAS\,02130+5509. The central map position and offsets are as in
Fig.\,1. } 

\label{fig:h2} 
\end{figure}

Both clouds display a striking cometary morphology with sharp
boundaries toward the N--NE, and much more diffuse tails extending
toward the S--SW. At the assumed distance of 2 kpc, the tails of both
clouds extend by about 4.5 pc. This kind of head-tail morphology is
typical of molecular clouds placed in the vicinity of massive stars;
winds and shocks from the massive stars are expected to compress the
nearmost edges of nearby clouds which thus develop the cometary
structure.

Both molecular clouds RNO\,6 and RNO\,6\,NW are also well observed 
in the $^{13}$CO  J=1$\rightarrow$0 map. Nevertheless, the CO and the
$^{13}$CO maps present  important differences which can be attributed
to several reasons. Most importantly, the CO J=2$\rightarrow$1
exhibit deep self-absorption features near the central velocities, at
which the $^{13}$CO profiles present their maxima. The differences in
opacities make the CO map very sensitive to variations in
kinetic temperature across the cloud, whereas
the  $^{13}$CO  J=1$\rightarrow$0 map is more
sensitive to variations in the gas column density and reveals the
opaque regions of both clouds. A well defined emission peak is
associated with  RNO\,6, and the main peak in the NW cloud is
placed near the offset ($-270${\arcsec}, +310{\arcsec}).  Finally,
the C$^{18}$O J=1$\rightarrow$0 emission is more concentrated than
the $^{13}$CO emission revealing that the most opaque zones are close
to offsets ($-20${\arcsec}, $-20${\arcsec}), (0, $-110${\arcsec}), 
and ($-270${\arcsec}, $+310${\arcsec}).

We have estimated the masses of the different clumps by assuming LTE
at a temperature of 15 K, which should be adequate for the inner
regions of the cloud.  We have considered that the boundaries 
of the clouds are defined by the lowest contours given in Fig.\,1 
(marked with dashed lines in the maps). 
From the C$^{18}$O data, by assuming a
C$^{18}$O/H$_2$ abundance ratio of 1.7$\times$10$^{-7}$ (Frerking et
al. 1982), we estimate that the three clumps near ($-20${\arcsec},
$-20${\arcsec}), (0, $-110${\arcsec}), and ($-270${\arcsec},
$+320${\arcsec}) have masses of 55, 34, and 73 $M_{\odot}$,
respectively. The masses of the RNO\,6 and RNO\,6\,NW molecular
clouds, estimated from the $^{13}$CO data are 190 and 235
$M_{\odot}$, respectively, where we have assumed a $^{13}$CO/H$_2$
abundance ratio of 2$\times$10$^{-6}$ (Frerking et al. \cite{frer82},
Bachiller \& Cernicharo \cite{bach86}). Thus, assuming that half of
the mass is in a halo which is  seen in CO but not in $^{13}$CO
(Cernicharo \& Gu\'elin \cite{cg87}), we estimate that the total mass
in both molecular clouds amounts to $\sim$850$M_{\odot}$. We caution
that these mass estimates are subject to important sources of
uncertainty. In addition to the poorly known distance, uncertainties
in the CO isotope abundances and in the excitation temperature make
the mass estimates to be uncertain by a factor of $\sim$2.

\subsection{Young stars in the region} 

\subsubsection{RNO\,6}

Figure~2 shows the new near-IR image of the RNO\,6 region overlayed 
on the Digital Sky Survey red image. The near-IR image, obtained
through a narrow-band H$_2$  filter, is shown here without
subtracting the continuum  in order to display the stars (which
disappear in the continuum-subtracted image). The central region,
which is  very bright in the optical, contains at least two stars
(labeled {\bf 1} and {\bf 2} in Fig.~2), one of which is the Herbig
AeBe star. This double source is accompanied by a  small cluster of
at least 10 fainter stars  which are distributed in a region of
$\sim$30{\arcsec}, in good agreement with Testi et al.
(\cite{tes98}). Only the most prominent stars in the central region
have been labeled. In addition to the stellar emission, the near-IR
image shows a kind of filament near offset (10\arcsec, 15\arcsec),
which is labeled {\bf F}, and another filamentary structure arising
in star {\bf 3}. The continuum-subtracted image of the area shows
that the emission from these filaments is purely H$_2$ line emission.
It can also be noted that the filament {\bf F} appears external to
the bipolar nebulosity and situated at the periphery of the optical
nebula.

\begin{figure}
\resizebox{\hsize}{!}{\includegraphics{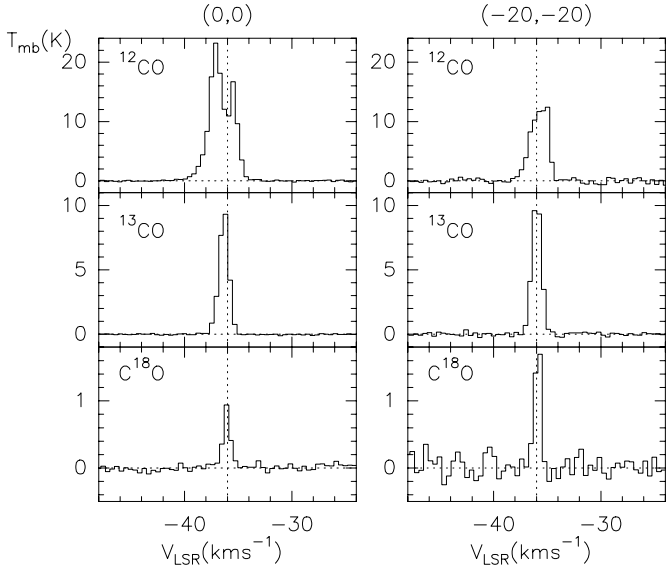}}

\caption{Spectra of CO $J=2{\ra}1$, $^{13}$CO $J=1{\ra}0$, and
C$^{18}$O $J=1{\ra}0$ observed toward position offsets (0,0) and
($-20{\arcsec}$, $-20{\arcsec}$) with respect to the nominal RNO\,6
central position (see Fig.\,1). The differences in the relative
intensities of the lines of CO and its isotopomers are discussed in the text.} 

\label{fig:spe_rno6} 
\end{figure}

We have examined the IRAS point source catalog to search for infrared
sources in the region. We found that RNO\,6 is coincident with an
IRAS source (IRAS\,02130+5509) of increasing spectrum from 12 to
100\,$\mu$m. The IRAS  fluxes (see Weaver \& Jones 1992, for
corrected co-added values) lead to a luminosity of
$\sim$300\,$L_{\odot}$.

The spectra of CO and its isotopomers arising from two positions near the
cluster are shown in Fig.~3. Note the striking differences  in the
emission of the CO isotopomers toward the (0,0) map position, i.e., the
position where the B star is located, and the (--20{\arcsec},
--20{\arcsec}) position where the C$^{18}$O is stronger. Clearly the
CO and C$^{18}$O lines peak at different places, and this is further
illustrated in the superposition of the near-IR image on the maps in
the lines of CO and its isotopomers shown in Fig.~4. The CO line
peaks near the central  (0,0) position, whereas the C$^{18}$O
maximum  is placed near (--20{\arcsec}, --20{\arcsec}). The strong CO
peak (peak line intensity $\sim$ 25 K) seen at the stellar position
clearly indicates the presence of  a local temperature maximum,
probably owing to the gas heated by the stellar radiation. In
contrast, the  C$^{18}$O maximum near position  (--20{\arcsec},
--20{\arcsec}) marks a maximum in gas column density in which the
temperature is significantly lower (peak CO line intensity $\sim$ 10
K).  These displacements  of the stellar positions from the dense
cores are typical of relatively ``evolved'' young stars (Fuente et
al. 2001), and indicate that  the UV flux from neighbouring stars to
the northeast has been able to destroy the molecules and to heat the
gas in this region. Finally, note  that the filament {\bf F} seen in
H$_2$ emission delineates the north-east edge of the C$^{18}$O clump.

\begin{figure*}
\vskip -5.75cm
\includegraphics{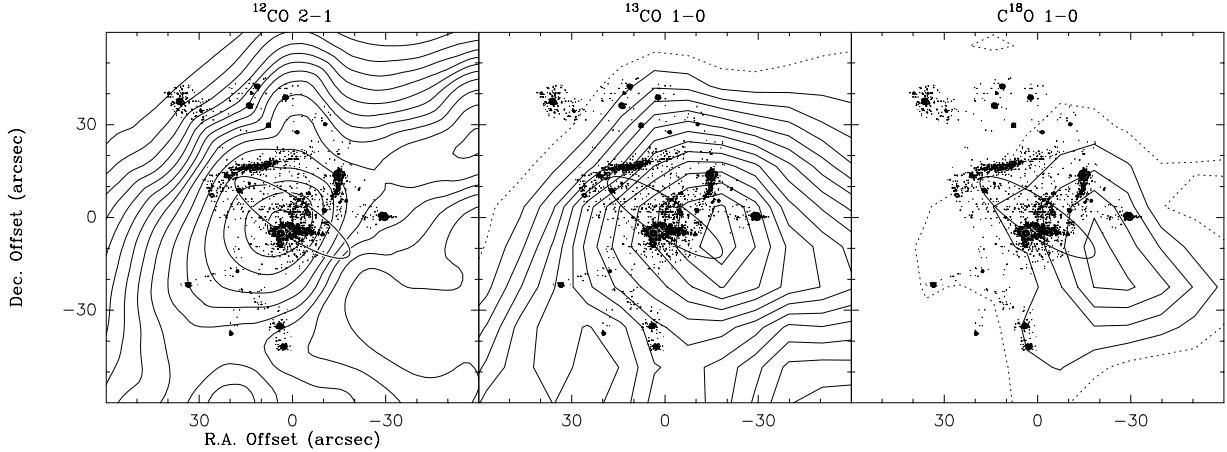}
\vskip 12cm

\caption{CO J=2$\rightarrow$1, $^{13}$CO J=1$\rightarrow$0, and
C$^{18}$O  J=1$\rightarrow$0 integrated intensity maps in the RNO\,6 
vicinity superimposed to the near infrared H$_2$ image. First contour
and step are 3 and 1.5 K\,\kms, for CO,  1.3 and 0.65 K\,\kms for
$^{13}$CO, and  0.42 and 0.21 K\,\kms for C$^{18}$O. The first contour
is drawn with dashed lines.  The ellipse indicates the position
uncertainty of IRAS\,02130+5509. The central map position and offsets
are as in Fig.\,1. } 

\label{fig:ir-co} 
\end{figure*}

\subsubsection{RNO\,6\,NW}

There is another point IRAS source, IRAS\,02124+5514, associated with
the maximum at ($-270${\arcsec}, $+320${\arcsec}) in RNO\,6\,NW.  This
source also presents an increasing spectrum from 12 to 100$\mu$m, but
it is much weaker than RNO\,6. The luminosity of IRAS\,02124+5514 is
estimated to $\geq$27\,$L_{\odot}$.

The near-IR images of the region we obtained with UKIRT are relatively
featureless. A detailed comparison with the Digital Sky Survey optical
images reveals a star with significant 2~$\mu$m excess emission at
position 02:12:29.1, 55:14:24 (1950.0), with $\sim$1$\arcsec$
uncertainty. This object is thus at the edge of the position
uncertainty ellipse of the IRAS source, so it is unclear whether the
IRAS and the near-IR sources are the same star.  On the other hand, in
the next section we report the detection of a high-velocity bipolar
outflow in this region. As shown in Sect. 4.2, the driving source of
the outflow is also predicted to be at the edge of the uncertainty
ellipse of the IRAS source, but it is $\sim$27$\arcsec$ east of the near-IR
source, so we believe that there are at least two young stars in this
region.

Figure 5 provides some molecular spectra observed toward RNO\,6\,NW.
H$^{13}$CO$^+$ J=1$\rightarrow$0 emission was detected around the
C$^{18}$O maximum, and the peak velocities and linewidths of both
lines are found to be similar, confirming that both lines are formed
in the same region of the cloud. The detection of H$^{13}$CO$^+$
implies the presence of rather high-density material (volume density
$n \geq$ a few 10$^4$\,cm$^{-3}$) associated with RNO\,6\,NW. The
presence of a dense molecular core, together with at least two
closely associated embedded sources demonstrates that the RNO\,6\,NW
molecular cloud is an active site of star formation.

\section{Kinematics}

\subsection{Velocity-position diagrams}

There are important variations in the CO profiles shapes and in the
central velocities of the different molecular lines across the mapped
region. This is illustrated in Fig.~6 by means of velocity-position
diagrams along the N-S direction for both RNO\,6 and RNO\,6\,NW
clouds.

The cuts near RNO\,6 show very nicely how the $^{13}$CO emission
lines peak at the  self-absorption dips or shoulders of the CO
profiles. The CO lines are broadened near the position of the IRAS
source, where both $^{13}$CO and C$^{18}$O  present well pronounced
maxima. Moving from this position to the South  (i.e. from the head to
the tail of the cometary cloud) there is a remarkable continuous
velocity gradient. The velocity of the peak emission varies from 
--36\,{\kms} to --34\,{\kms}.

A similar cut near RNO\,6\,NW displays a rather similar behavior. 
Again a systematic velocity gradient is observed from the  head to
the tail of the cloud, with the velocity  changing from --36\,{\kms}
near the head position  to --37\,{\kms} at the end of the tail.  The
CO lines are also found to be very broad around the position of  the
IRAS source. Nevertheless, the broadening around IRAS\,02124+5514 is
much more pronounced than that observed around RNO\,6 as the CO lines
extend over $\sim$10\,{\kms} and exhibit prominent wings typical of 
high-velocity molecular outflows.

\subsection{A bipolar outflow around IRAS\,02124+5514 (RNO\,6\,NW)}
 
The CO spectra around IRAS\,02124+5514 (see Fig. 5) show that the
wings extend from about --41 {\kms} to --30 {\kms}. Examining the
$^{12}$CO J=2$\rightarrow$1  profiles around IRAS\,02124+5514 we find
that there is a systematic behavior in the spatial location  of the
velocity wings. Blue-shifted wings are observed at the positions
north of the IRAS source, whereas redshifted wings are observed at
the southern positions. The distribution of the blueshifted wing
(integrated from --41 to --37.5 {\kms}) and of the redshifted wing
(integrated from --34.5 to --30 {\kms}) is shown in Fig. 7. The
distribution  of the emission is strongly bipolar with respect to the
position offset (--265\arcsec, +315\arcsec). We thus predict that the
source driving the outflow is at the (corresponding) coordinates:
02:12:32.1,  55:14:27 (1950.0), with a uncertainty of $\pm$5$\arcsec$.
As discussed in Sect. 3.2.2, this position is at the edge of the error
position ellipse of the IRAS source, but it is $\sim27\arcsec$ east from
the near-IR source we detected in the region.  The three positions are
marked in the outflow map of Fig.\,7 with different symbols. If we
take into account the resolution of our observations and the IRAS
positioning error, we conclude that the IRAS\,02124+5514 emission
could arise in the outflow driving source, in the near-IR star, or in
both of them. However, the large position offset from the outflow
driving source to the near-IR star suggests that these are different
objects.

\begin{figure}
\resizebox{\hsize}{!}{\includegraphics{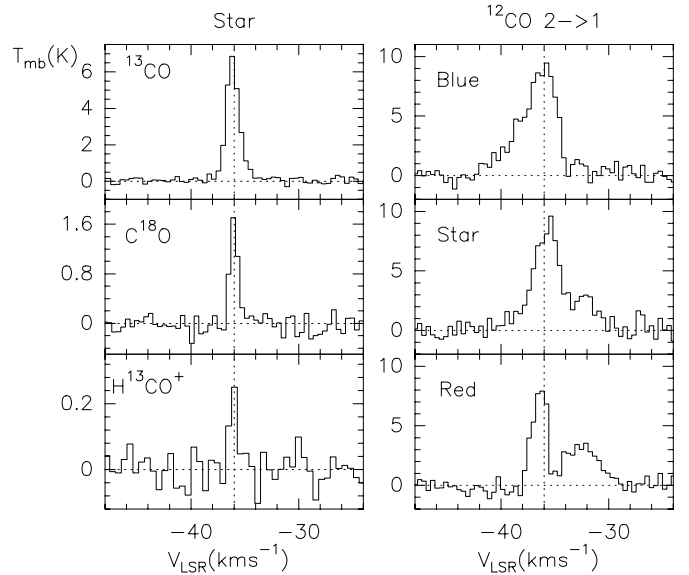}}

\caption{Some spectra observed around RNO\,6\,NW.  Left.- $^{13}$CO,
C$^{18}$O, and H$^{13}$CO$^+$ $J=1\ra0$ spectra observed toward the
position of the C$^{18}$O maximum at ($-265\arcsec$, $+315\arcsec$).
Right.- $^{12}$CO $J=2\ra1$ spectra observed toward the C$^{18}$O peak
position (central panel).  The two spectra labeled ``Blue'' and
``Red'' correspond to positions ($-260\arcsec$, $+330\arcsec$) and
($-260\arcsec$, $+260\arcsec$), which are placed north and south from
the C$^{18}$O peak, respectively.  These profiles provide evidence
for a bipolar outflow in RNO\,6\,NW. }

\label{fig:spe} 
\end{figure}

The total length of the outflow is 0.9~pc, and it is poorly
collimated. Its collimation factor, estimated as the ratio of the
outflow length to its width, is $\sim 2$. The outflow is very
asymmetric; the south (redshifted) lobe is brighter, and twice more
extended along the  flow axis than the north (blueshifted) lobe. Such
an asymmetry is likely related to the location of the IRAS source
within the ambient molecular cloud. It is important to note here that
molecular outflows are made of accelerated ``ambient'' material (e.g.
Bachiller \& Tafalla \cite{bt99}). Since IRAS\,02124+5514 is near the
north border of the globule's head, the southern  flow lobe is
propagating within the cloud, where there is abundant material to be
swept up. However the north outflow lobe is propagating out of the
cloud into an ``empty'' medium which is illuminated by a strong UV
field (see Sect.~5). So in this  northern area there is less
material to be swept up, and moreover the UV field can quickly
destroy the CO molecules carried out of the cloud. We thus believe
that the cloud morphology, and the location of the YSO within it, can
explain the observed asymmetry of this outflow. Indeed there are
other known cases of asymmetric outflows at the edges of molecular
clouds, examples include HH\,46-47 (Hartigan et al. \cite{ha90}) and
Orion\,B (Richer et al. \cite{rich92}).

The mass of the outflow, $M$, can be estimated from the CO
intensities integrated to the wings. We assumed Local Thermodynamical
Equilibrium (LTE) at  an excitation temperature of 15\,K, optically
thin emission along the wings, and a H$_2$/CO abundance ratio of
1$\times$10$^{-4}$ (Frerking et al. \cite{frer82}). We found that the
masses of the blue and redshifted lobes are 0.12 and 0.35
$M_{\odot}$, respectively. We however caution that this mass estimate
is subject to important uncertainties. In addition to that arising
from the poorly known distance, the main uncertainties  arise (i) in
the placement of the velocity boundary between the high velocity wing
and the  ambient line and (ii) in the assumed kinetic temperature
(note that the mass estimate is nearly proportional to the assumed
temperature). We believe that our estimate of the outflow mass ($M
\sim$ 0.5 M$_{\odot}$) is accurate within a factor of 2.

The kinematical time scale of the outflow can be estimated, for each
of the lobes, as the ratio $\tau \sim R/V_{mean}$ where $R$ is the
distance from the center of mass of the lobe to the driving source,
and $V_{mean}$ is the mean velocity of the lobe. We estimate that
both outflow lobes have $\tau \sim$ 10$^5$ yr. This estimate
considers that  the outflow is created from a unique explosive
episode, which very likely does not corresponds to reality. As
mentioned above, the CO outflow consists  of ambient material which
is accelerated from the driving agent (the primary wind from the
star/disk system) and which is decelerated as it moves into the
surrounding molecular cloud. So the estimate  given above is obtained
from a very crude approximation. Another important source of
uncertainty in estimating the kinematic time scale is the inclination
angle of the outflow with respect to the line of sight. The estimate
above is valid for an inclination angle $i$ = 45$\deg$, otherwise it
has to be multiplied by a factor of tan($i$). The small spatial 
overlap between the blueshifted and redshifted lobes indicates that
the inclination angle of the RNO\,6\,NW outflow is probably higher
than 45$\deg$.  If the inclination were $i$ = 75$\deg$, the estimated
time scale should be multiplied by a factor 0.27.

The momentum, kinetic energy, and mechanical power of the outflow can
be estimated as $M\,V_{mean}$, $M\,V{^2}_{mean}/2$ and 
$M\,V{^2}_{mean}/(2\,\tau)$, respectively. We obtain 1.4
M$_{\odot}${\kms}, 4.2$\times$10$^{43}$ erg, and  5$\times$10$^{-3}$
L$_{\odot}$, respectively, without correcting for projection effects.
For an inclination angle of $i$ = 75$\deg$, these estimates become
5.4 M$_{\odot}${\kms}, 6.4$\times$10$^{44}$ erg, and  0.26
L$_{\odot}$, respectively. Thus, compared to other outflows, the
parameters of the RNO\,6\,NW outflow appear relatively modest and
correspond to a driving young star of low mass. The mechanical power
of the outflow, in the range 5$\times$10$^{-3}$ to 0.26 L$_{\odot}$,
depending on the outflow inclination, points to a source of 1-- 20
L$_{\odot}$ (see e.g. Bachiller \& Tafalla \cite{bt99}).
By comparing this estimate with the luminosity of the IRAS\,02124+5514 source 
($\geq$27\,$L_{\odot}$), we conclude that the IRAS flux could
contain some contribution from the near-IR star placed $\sim$27{\arcsec} west 
from the outflow origin (see Fig. 7).

Since SiO emission is a sign of youth in outflows (Bachiller
\cite{ba96}), we searched for SiO emission in several selected
positions of the RNO\,6\,NW outflow to assess the evolutionary stage
of the system. No SiO emission was detected at a level of 0.02\,K
rms.

In summary, the rather long kinematic time scale ($\sim$ 10$^5$ yr), 
poor collimation ($\sim 2$), 
and lack of SiO emission indicate that the IRAS\,02124+5514 outflow is in a
rather evolved stage. These outflow characteristics  (see Bachiller
\& Tafalla \cite{bt99}) indicate that IRAS\,02124+5514 could be a
relatively evolved Class I source with a luminosity in the range
1-- 20  L$_{\odot}$.

\section{Discussion}

The cometary morphology of both RNO\,6 and RNO\,6\,NW molecular
clouds, the presence of star formation activity at the dense heads, 
and the strong velocity gradients observed along the major axes of
both clouds, are all attributes of molecular clouds which are subject
to the strong influence of neighboring luminous stars.

\begin{figure}
\resizebox{\hsize}{!}{\includegraphics{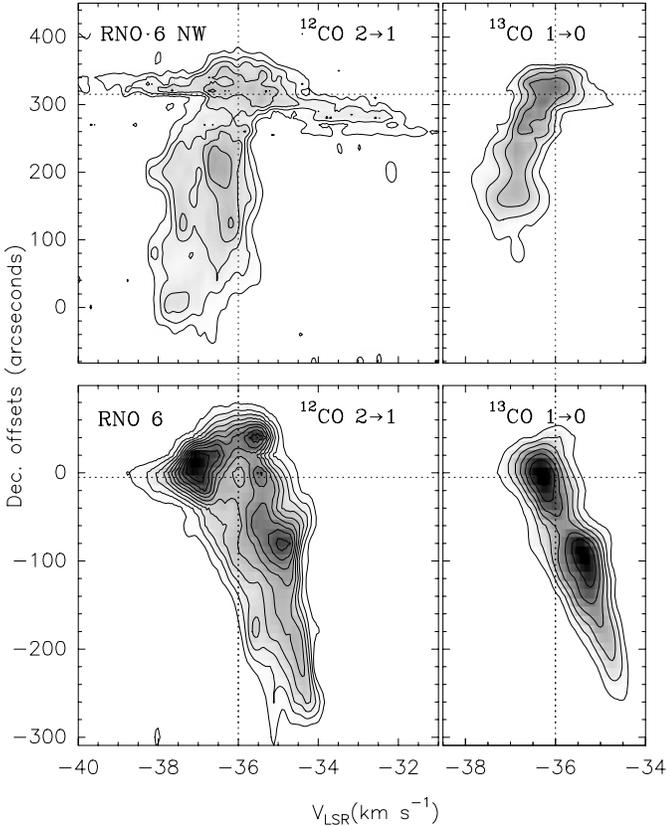}}

\caption{Velocity-position diagrams along lines at constant right
ascension for the $^{12}$CO J=2$\rightarrow$1 and $^{13}$CO
J=1$\rightarrow$0 line emission. The diagrams in the bottom panels
correspond to the RNO\,6 cloud, and are made along a line passing on
the HAeBe star  (right ascension offset = 0, in the maps of Fig. 1).
The position of the HAeBe star is indicated with an horizontal dotted
line. The diagrams in the upper panels correspond to the RNO\,6\,NW
cloud, and are made along a line passing on the center of the high 
velocity outflow. Note the broad wing emission from the outflow at
the globule head. The horizontal dotted line indicates the outflow
center.  First contour levels and steps are 2.5~K for $^{12}$CO and 
1.5~K for $^{13}$CO.    In all panels a vertical dotted line at --36
km\,s$^{-1}$ (i.e. close to the velocity of the quiescent material at
the heads of the globules) is marked for orientation.  } 

\label{fig:cortes} 
\end{figure}

Such category of cometary clouds include the bright-rimmed globules
as those studied by Hawarden \& Brand (\cite{hb76}) and Sugitani et
al. (\cite{sug89}). Cometary globules are believed to be formed by
the Radiation-Driven-Implosion (RDI) mechanism first described by
Reipurth (\cite{br83}), and later on modeled by Bertoldi \& McKee
(\cite{bm90}) and Lefloch \& Lazareff (\cite{ll94},\cite{ll95}). In
this mechanism,  the incident photons from neighboring bright stars
ionize the gas at the globule surface which thus begin to flow out. 
Moreover, the high pressure at the globule surface drives a shock
wave into the globule which compresses the molecular gas.  Such
compression can lead to the formation of new stars at the globule
head, whereas the external layers of the cloud continue being
photo-evaporated. As a result of this mechanism, as first noted by
Sugitani et al. (1989), the ratio of the stellar luminosity to the
mass of the cloud is found to be much higher 
(0.3-- 13 $L_{\odot}/M_{\odot}$) in
cometary clouds than in nearby isolated dark clouds (0.03-- 0.3
$L_{\odot}/M_{\odot}$). For the RNO\,6 and RNO\,6\,NW clouds, the
ratio is $\sim$3 and $>$0.3, respectively, which confirms that star
formation has been triggered by the action of some external luminous
stars in RNO\,6, and very probably in RNO\,6\,NW.

The cometary structures of both RNO\,6 and RNO\,6\,NW clouds are
elongated north-south with the heads oriented to the north. 
Moreover, the northern edge of the RNO\,6 cloud exhibits bright H$_2$
emission (feature {\bf F} in Fig. 2), suggesting that a shock wave
could be propagating from north to south. We thus explored the region
north of these clouds for the presence of massive stars, and we found
that the double cluster $h$ and $\chi$ Persei (NGC\,869 and NGC\,884)
lies $\sim$1.5{\deg} away from the clouds. The double cluster  is
believed to be the core of the large OB association Per\,OB1 (see
e.g. Cappa \& Herbstmeier, \cite{ca00}, and the references therein).
Fig.~8 shows all the known OB stars in a 2$\deg$ region to the north
of RNO\,6 clouds. O stars are denoted by big filled stars, B3 or
earlier B stars by medium filled stars, and B stars later than B3 by
small filled stars.  White (unfilled) stars represent B stars for 
which the exact spectral class is unknown. Clearly these stars can be
responsible for an intense UV field at the heads of the RNO\,6 and
RNO\,6\,NW clouds.

We estimated the ionizing UV photon flux at the cometary heads of the
RNO\,6 and RNO\,6\,NW clouds due to the earliest type stars in the
region -- including those marked with their names and spectral types 
in Fig.~8--
by using the stellar properties listed by Panagia (\cite{pa73}), and
by assuming that all stars and both molecular clouds are in the same
plane perpendicular to the line of sight. We obtain that the Ly-c flux 
incident on the cloud heads is $\sim$6$\times$10$^{8}$
cm$^{-2}$s$^{-1}$. But it should
be noted that this is a very rough estimate of the actual UV flux
because of several reasons.  First there are many B stars whose exact
sub-classification is unknown, and every B0--B1 type star can
contribute up to a few percent of the total flux. Second, very distant stars
out from the field covered by Fig.\,8 can have a significant
contribution if they are of early types.  Third, we have assumed all
stars in the same plane which is an obvious simplification. In any
event, we believe this flux to be accurate within a factor of 5.
As we next discuss, this flux of UV photons is large enough to cause
the observed cometary structures.

\begin{figure}
\resizebox{\hsize}{!}{\includegraphics{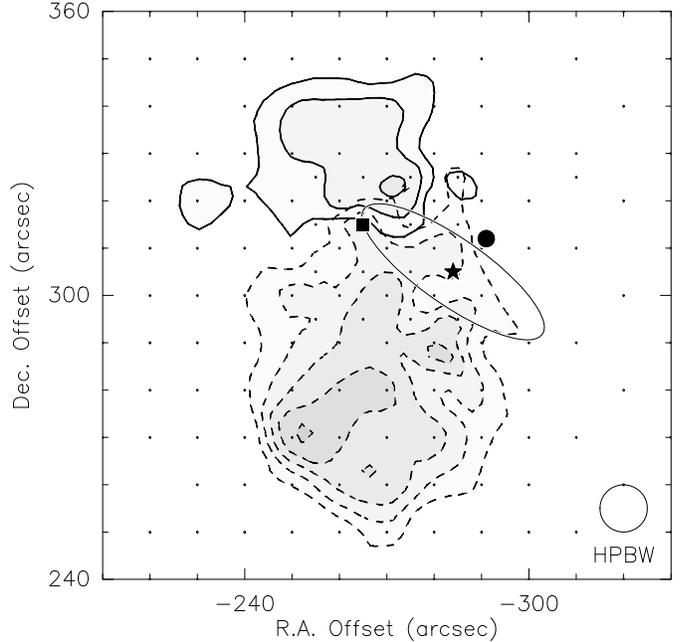}}

\caption[]{Integrated emission of CO $J=2\ra1$ showing the bipolar
structure of the outflow around IRAS\,02124+5514, in the RNO\,6\,NW
region.  Velocity intervals are from --41 to --37.5 \kms\ for the
northern (blueshifted) lobe and from --34.5 to --30 \kms\ for the
southern (redshifted) lobe.  First contour and contour spacing are 6
and 3.5 K \kms, respectively.  Solid contours are for the blueshifted
gas, whereas dashed contours are for the redshifted gas.  The star
and the ellipse mark the nominal position of the IRAS source and its
error uncertainty, the filled circle the position of a near-IR star
(see text), and the square the expected position for the outflow
driving source.  Position offsets are with respect to the same
central position of Fig.\,1.}

\label{fig:2}
\end{figure}

Since the RDI mechanism is well documented in the literature, we
looked for the best RDI model that would match the observed 
morphology and kinematics of the RNO\,6 and RNO\,6\,NW clouds.  We
found that model 2 of Lefloch and Lazareff (\cite{ll94}, hereafter
LL94) is in striking agreement with our observations.  In fact,
several aspects of  the morphology and kinematics of both RNO\,6 and
RNO\,6\,NW clouds are in very good agreement with the structure of a
cloud at the re-expansion phase described in model 2 of LL94:

\noindent
1.- The general appearance of both RNO\,6 and RNO\,6NW
molecular clouds is very similar 
to that of the model 2 globule at the re-expansion phase.

\noindent
2.- There is a clear velocity gradient in the north-south direction
of  both RNO\,6 and RNO\,6NW clouds (as seen in Fig.~6, and described
in Sect.~4.1).  In model 2 (LL94), such gradients are caused by the 
acceleration of the gas along the tail of the globules. 

\noindent
3.- Line broadening can be seen at the heads of both molecular clouds
(see Fig.~6). In model 2 (LL94), such a broadening is indeed
predicted, it results from the expansion of the forward and backward 
sides of the globule's head.  

\noindent
4.- The position of the CO peak is situated about 60$\arcsec$ below
the surface of the cometary bright rim of these clouds. This maximum
density point is not immediately behind the bright rim (as expected
for the quasi-stationary state) but further behind.

Although the model predictions of LL94 are in the context of small
cometary globules it has been suggested that the model can be applied
to a different case with an appropriate scaling of the parameters.
The corresponding scaling factor is denoted by $k$. The meaning of this
factor is such that in order to match the parameters of any cloud with the
model,  the size and the age of the cloud described in model 2 (LL94) 
have to be multiplied by a factor of $k$, the mass by a factor 
$k^2$, and the cloud density and the incident Ly-c flux
by a factor $k^{-1}$. 
For the case of the molecular clouds discussed here, the matching of the
sizes and masses of the RNO\,6 and RNO\,6\,NW clouds with those of the
model leads to a scaling factor $k\sim$4. 
Then we find that the required
Ly-c flux incident on the clouds surface is 5.5$\times10^8$
cm$^{-2}$s$^{-1}$, in  general agreement with the estimate made from
the fluxes of the neighboring stars. The age resulting from the model, i.e.
the duration of the photo-ionization phase, is 1.3\,Myr.

Further comparison of the observations and the model provides
additional pieces of information. For instance,  the sign of the
head-tail velocity gradient can be used to infer the position in
space of the cometary clouds.  The red-shifted gradient in RNO\,6
indicates that the tail is pointed towards us, whereas the
blue-shifted gradient in RNO\,6NW indicates that it is pointing away
from us.  The observed magnitude of the velocity gradient $\Delta v$
corresponds  to a kinematical timescale of t$_{kin}$$\sim
\frac{1}{\Delta v} \sim$1.4~Myr, in very good agreement with the 
value of 1.3\,Myr obtained for the duration of the photo-ionization phase.
This timescale is also consistent with the ages of the $h$ and $\chi$ 
stellar clusters, which are of the order of 6 and 3\,Myr respectively,
according to Tapia et al. (1984).
We thus believe that 1.3\,Myr can be considered as a good
estimate of the time spent by these clouds under the incidence of
strong UV  flux.

Additional support for triggered star formation in the RNO\,6 cloud
comes from the distribution of young stars in the cloud's head. In
Fig.~4 we can see that the distribution of the stars are elongated or
fan out towards the north away from the IRAS source. The only star
with H${\alpha}$ emission in this cluster is the RNO\,6 object itself
which is the southernmost member of the cluster.  The spatial
coincidence of the IRAS source with the H${\alpha}$ emission star
suggests the relative youth of this star with respect to other
cluster members. Also, the IRAS source and the two stars with
associated nebulous filaments are situated inside the C$^{18}$O
contours that represent the densest portion of the cloud, which
confirms the relative youth of these near-IR sources.  The placement
of bluer stars closer to the surface of the cloud's head  and redder
stars away from the cometary head, but closer to the  IRAS source and
C$^{18}$O maximum, is a strong indication of triggered sequential star
formation (see also Sugitani et al. \cite{sug95}).

\begin{figure}

\vskip -2cm
\resizebox{\hsize}{!}{\includegraphics{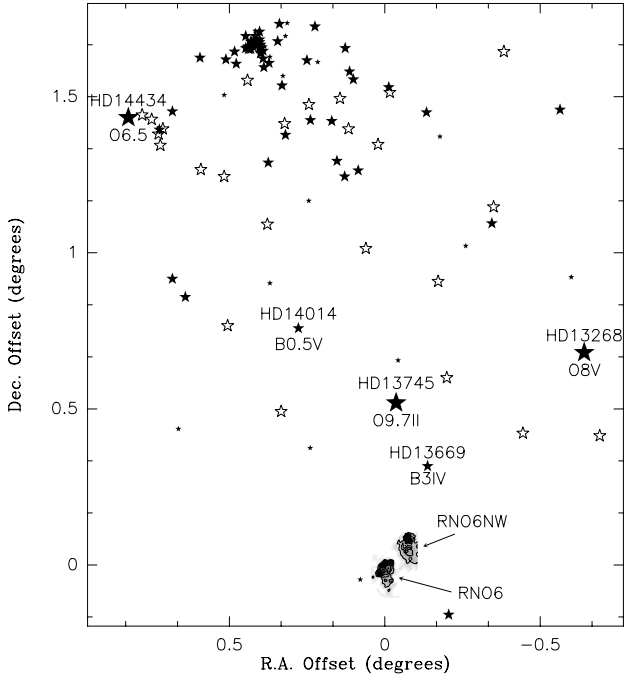}}
\caption{OB stars to the north of the RNO\,6 and RNO\,6\,NW molecular clouds.
Positions and spectral types are from the CDS Simbad database, and completed 
with the compilation of Cappa \& Herbstmeier (\cite{ca00}). 
O stars are denoted by big filled stars,
B3 or earlier B stars by medium filled stars, and B stars later than
B3 by small filled stars.  Unfilled stars represent B stars for which the
exact spectral class is unknown. Some stars of known spectral types
that may illuminate the RNO6 clouds with significant UV flux are marked with 
their names. The large stellar concentration 1.5$\deg$ to the north 
is the double cluster $h$ and $\chi$ Persei, which is believed to be
the core of the OB association Per\,OB1.}
\label{fig:8}
\end{figure}

Many cometary globules have been observed until now in different
evolutionary stages.  Cometary globules that also agree with   model
2 of LL94 have been found near the Rosette nebula (Patel et al.
\cite{patel93}, White et al. \cite{white97}) and near the Gum Nebula
(Gonz\'alez-Alfonso et al \cite{alf95}). Some of these globules
harbor stars and some times bipolar outflows  (Nielsen et al.
\cite{niel98}, Cernicharo et al. \cite{cer92},  Codella et al.
\cite{cod01}). However, in all cases mentioned above the globules are
low mass objects (up to tens of solar masses) containing stars of
masses around 1\,M$_{\sun}$. In contrast, the RNO\,6 and RNO\,6NW
clouds, with $\sim$200\,M$_{\sun}$ each (as estimated from
$^{13}$CO), are higher mass counterparts to these cometary globules
studied before. A precedent to such high mass objects is the
135\,M$_{\sun}$ globule studied by Lefloch et al. (\cite{le97}) in
IC\,1848. It thus appears that the RDI mechanism is able, not only to
form low-mass stars in small globules, but also to form intermediate
mass stars and small clusters in massive molecular  clouds.

\section{Conclusions} 

We have presented mm-wave line observations of the molecular clouds
around RNO\,6. The main results of this work can be summarized as
follows:

\begin{itemize}

\item
We have mapped the molecular cloud harboring RNO\,6 along with
a newly detected molecular cloud $6'$ northwest
of RNO\,6 which has been named RNO\,6\,NW. 

\item
These clouds RNO\,6 and  RNO\,6\,NW display striking similarities in
their cometary structures and overall kinematics.  By using $^{13}$CO
line observations, we estimate that both clouds have  similar sizes
($\sim$4.5\,pc) and masses ($\sim$200 M$_{\sun}$).

\item
Both molecular clouds RNO\,6 and RNO\,6\,NW are active in star
formation. From new near-IR narrowband images, we confirm that RNO\,6
hosts an embedded IR cluster that includes a Herbig Be star. A
conspicuous H$_2$ filament is found to delineate the dense cometary
head of the globule.  

\item
RNO\,6\,NW hosts at least two IR sources and a bipolar molecular
outflow $\sim$0.9\,pc of length and $\sim$0.5\,M$_{\sun}$ of mass.

\item
The cometary structure of both clouds RNO\,6 and RNO\,6\,NW has been
created by the UV radiation from numerous OB stars lying 1.5{\deg} to
the north.  Such OB stars are associated with the double cluster $h$
and $\chi$ Persei, and are probably members of the Per\,OB1
association.

\item
Star formation inside these molecular clouds has been very likely
triggered by the Radiation Driven Implosion (RDI) mechanism. From
comparison with RDI theoretical models, we find that the similar
kinematics and morphology of both clouds is well explained if they
are at a re-expansion phase.  

\item
Triggered sequential star formation also explains the observed
spatial distribution of the members of the near-IR cluster inside the
RNO\,6 cloud, and the morphology of the H$_2$ filament. 

\item
The RNO\,6 and RNO\,6\,NW clouds are high-mass counterparts to the
cometary globules of smaller masses which have been studied up to
now. Thus our observations demonstrate that the RDI mechanism can
produce, not only  low mass stars in small globules, but also
intermediate mass stars and clusters in massive clouds.

\end{itemize}

\acknowledgements 

{The authors are grateful  to Dr. M. P\'erez Guti\'errez for help
with the mm-wave observations, to Dr. C. J. Davis for help with the
near-IR observations, to Dr. B. Lefloch for useful discussions, and
to the referee, Dr. W.H. McCutcheon, for careful reading the manuscript
and for valuable comments and suggestions.
The SIMBAD database is operated by the CDS, Strasbourg, France.
The United Kingdom Infrared Telescope is operated by the
Joint Astronomy Centre on behalf of the U.K. Particle Physics and
Astronomy Research Council. This research have been partially
supported by Spanish DGES grant AYA2000-927.}

\end{document}